# Tensile properties of all-polymeric syntactic foam composites: experimental characterization and mathematical modelling

Zeshan Yousaf[1, 2*], Neil F. Morrison[1], William J. Parnell[1]

[1]Department of Mathematics, University of Manchester, Oxford Rd, Manchester M13 9PL, UK

[2]Department of Materials, University of Manchester, Manchester M13 9PL, UK

* zeshan.yousaf@manchester.ac.uk

**Abstract**

All-polymer syntactic foams are studied under large strain cyclic and monotonic tensile loading in order to reveal their tensile stress-strain behaviour, recoverability, tensile strength, and elongation at break. The syntactic foam under study here consists of hollow thermoplastic microspheres (HTMs) of two distinct grades (551 and 920), with distributions of mean-wall thicknesses and diameters, embedded inside a polyurethane matrix in various volume fractions. Cyclic loading-unloading curves are recorded, revealing the level of viscoelasticity exhibited by the materials (which becomes a stronger effect with increasing volume fractions of HTMs) and indicating the level of repeatability of loading under large strain. Samples are also subjected to monotonic tensile loading in order to study their elongation at break. Higher volume fractions of HTMs increase the stiffness of the material and whilst it is observed that the materials are highly elastic over a wide range of tensile strains, damage arises at lower levels of strain for more highly filled materials. The HTM syntactic foams thus exhibit lower breaking strains compared to the neat matrix, which is attributed to matrix-microsphere interfacial debonding. Furthermore, by employing optimization techniques, linear elastic properties of the microspheres and an average shell thickness of the 551 grade are inferred by comparing experimental results to predictions from the Generalized Self-Consistent Method, incorporating polydispersity data on the size distribution of the microspheres. These results complement previous work which involved direct experimental measurements of the 920 grade shell thickness. Results also indicate that the characterization of microsphere properties is not critically dependent on access to high resolution microsphere diameter distribution data, provided that an accurate representative mean diameter is known. Finally, the thermal degradation of the samples is studied by using thermogravimetric analysis (TGA).

## 1. Introduction

Syntactic foams (SFs) are lightweight composite materials that are employed in e.g. the aerospace, automotive, and subsea sectors due to their excellent mechanical and damage tolerance properties, high specific strength, and low water absorbency [1-4]. SFs are manufactured by adding hollow thin-walled particles, known as microspheres or microballoons, into the matrix material. The effective mechanical properties of a SF can be tailored by selecting an appropriate combination of matrix material and hollow microspheres [5]. It has been observed that the wall-thickness and diameter of the microspheres plays an important role in determining the mechanical properties of SFs [3, 6-9]. SFs have been manufactured comprising a range of hollow particles (e.g. glass, ceramic, and polymers) and matrix materials [10-17]. SFs containing hollow glass microspheres (HGMs) are the most popular and prevalent for load-bearing applications due to their high specific strength and modulus. The response of these HGM SFs has been investigated thoroughly under compression, tension, and shear loading [2, 3, 7, 11, 14, 18-23].

It has been revealed recently that a new class of all-polymer SFs, manufactured by incorporating hollow thermoplastic microspheres (HTMs) into a polyurethane matrix, exhibit strong recoverability, damage tolerance, and energy dissipation under cyclic compression [17] and in particular they can be subjected to high compressive strains repeatedly. However, no literature is available on the *tensile* properties of these HTM SFs when subjected to large strains. A thorough investigation of these HTM SFs is required under various loading conditions to establish an improved understanding of their mechanical behaviour, and to develop associated simulation tools. In the present work, we therefore study these HTM SFs under cyclic and monotonic tensile loading to reveal their stress-strain relationship, recoverability, damage tolerance, and elongation at break. For this purpose, SFs are manufactured by incorporating two different grades of HTMs (with different mean wall-thicknesses and diameters) into a polyurethane matrix at various volume fractions. The effect of mean wall-thickness, diameter, and volume fraction of HTMs on the tensile properties of the SFs is studied. The SFs are subjected to cyclic tensile loading and the associated stress-strain curves are recorded. Loading-unloading curves are obtained for both virgin (untested) samples and samples that had been previously tested, in order to study the tensile behaviour and recoverability of these materials. Monotonic tensile tests are also carried out to study the elongation of the SFs at break. Furthermore, we consider the physical characterization of the HTMs in question, by using a model based on the Generalized Self-Consistent Method, combining geometrical polydispersity data obtained via imaging techniques [24] with the measured tensile response of HTM SFs. By calculating the model's prediction of the stiffness of a SF containing HTMs with sizes and aspect ratios distributed identically to the imaging data, we are able to solve iteratively for the optimal fitting values of the unknown (or unverified) properties of the HTMs. In particular, we extend previous results on the characterization of the HTMs by computing values for their Young's moduli and Poisson's ratios, comparing these to analogous values found in [24] via fits to compression experiments [4], and we determine a representative value of the shell thickness of 551 DE 40 d42 grade HTMs, complementing the directly measured value of the average shell thickness of the 920 DE 80 d30 grade HTMs determined in [24]. Furthermore, the thermal degradation of these HTM SFs is studied by using thermogravimetric analysis (TGA).

## 2. Material and mechanical testing

*2.1 Material*

HTM SFs were fabricated by introducing HTMs into a polyurethane elastomeric matrix. Hollow polymer microspheres were incorporated into the matrix at a range of representative volume fractions: 2%, 10%, and 40%. The two grades of HTMs considered here are 920 DE 80 d30 and 551 DE 40 d42 (Expancel grades supplied by Nouryon), referred to below as "920" and "551" for brevity. Details of the 551 and 920 grades, which have different mean wall-thicknesses and diameters, are provided in Table 1. Scanning electron microscope (SEM) images of the HTMs and HTM SFs are presented in Figures 1 & 2 respectively. Measured and theoretically calculated densities of the resulting HTM SFs are presented in Table 2. The theoretical densities are obtained using the rule of mixtures [3]. Values for both the measured and theoretical densities are very similar, suggesting a minimal destruction of HTMs during mixing and minimal presence of voids.

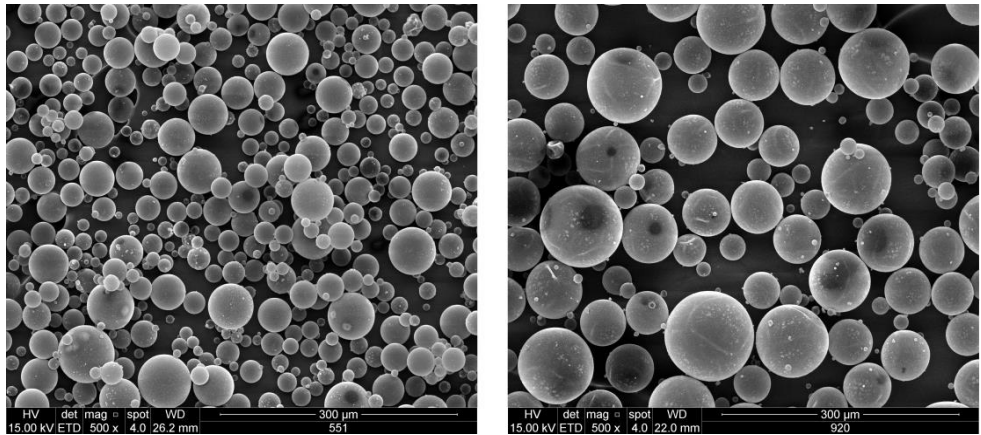

**Fig. 1** SEM images of HTM grades 551 (left) and 920-40% (right)

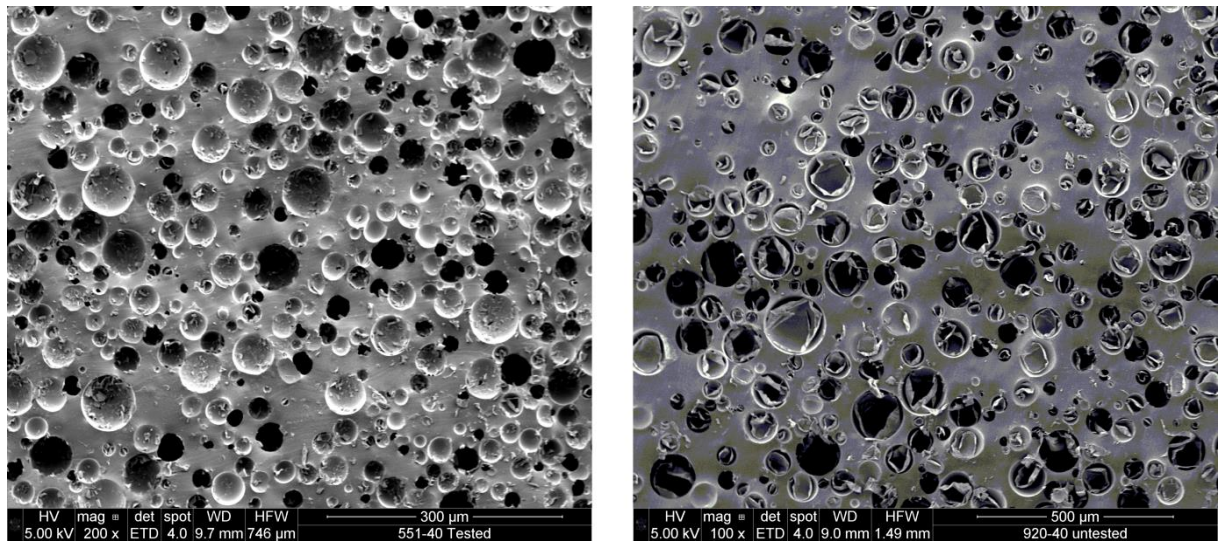

**Fig. 2** SEM images of SFs containing HTMs of grade 551-40% (left) and 920-40% (right).

**Table 1** Details of HTMs (551 DE 40 d42 & 920 DE 80 d30).

| Microsphere grade | Microsphere diameter (micron) - by manufacturer | Shell thickness (micron) - by manufacturer | Density (g/cm$^3$) | Wall thickness-to-diameter ratio | Volume-weighted mean diameter (micron) [24] | Shell thickness (micron) [24] |
|---|---|---|---|---|---|---|
| 551 | 40 | 0.25 | 0.042 | 0.00625 | 81.563 | |
| 920 | 70 | 0.35 | 0.030 | 0.00500 | 39.544 | 0.29 |

**Table 2** Measured and theoretical densities of HTM SFs.

| Material type | Unfilled | 551-2% | 920-2% | 551-10% | 920-10% | 551-40% | 920-40% |
|---|---|---|---|---|---|---|---|
| Measured density (g/cm$^3$) | 1.081 | 1.060 | 1.061 | 0.975 | 0.985 | 0.666 | 0.673 |
| Theoretical density (g/cm$^3$) | n/a | 1.060 | 1.06 | 0.977 | 0.976 | 0.665 | 0.661 |

## 2.2 *Mechanical testing*

Uniaxial cyclic tensile tests were performed on the HTM SF samples using an Instron universal testing machine equipped with a 100 kN load-cell. The BS ISO 37:2017 standard for tensile testing of vulcanized rubber is followed. The samples were machined to a dumb-bell shape (Fig. 3) (sample dimensions were maintained according to type 1 of BS ISO 37:2017). They were subjected to cyclic tensile loading at a cross-head speed of 500 mm/min. Five subsequent successive loading and unloading curves to a specified strain level were recorded. Monotonic tensile tests to failure were also carried out at the same cross-head speed as the speed used for cyclic tests (500 mm/min).

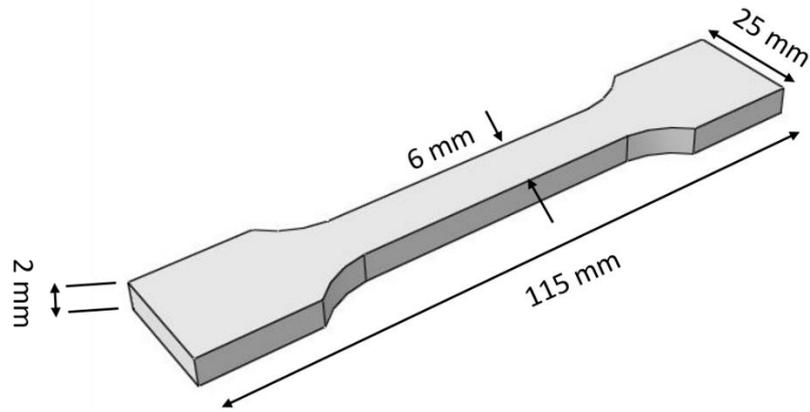

**Fig. 3** Geometry of the dumb-bell test piece, according to BS ISO 37:2017.

### 3. Results and discussion

3.1 Macroscopic deformation under cyclic loading

Cyclic uniaxial tensile testing was carried out on all samples to 25% and 50% strains successively, as depicted in Fig. 4. Initially, cyclic loading was applied to virgin samples up to 25% strain. Then, after one week, cyclic testing was repeated on these same samples to the same strain level. Thereafter, a similar procedure was adopted for the samples up to 50% strain.

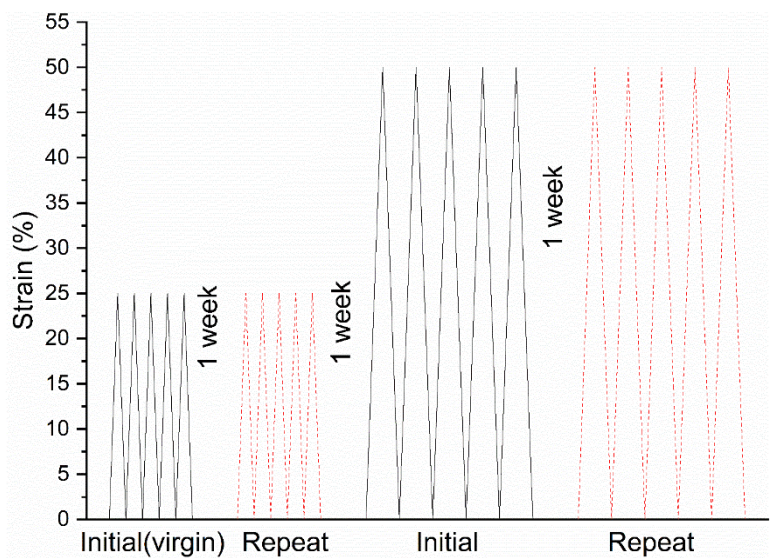

**Fig. 4** Pattern adopted for cyclic tensile loading, with initial testing represented by solid lines whilst repeated tests are represented by dotted lines.

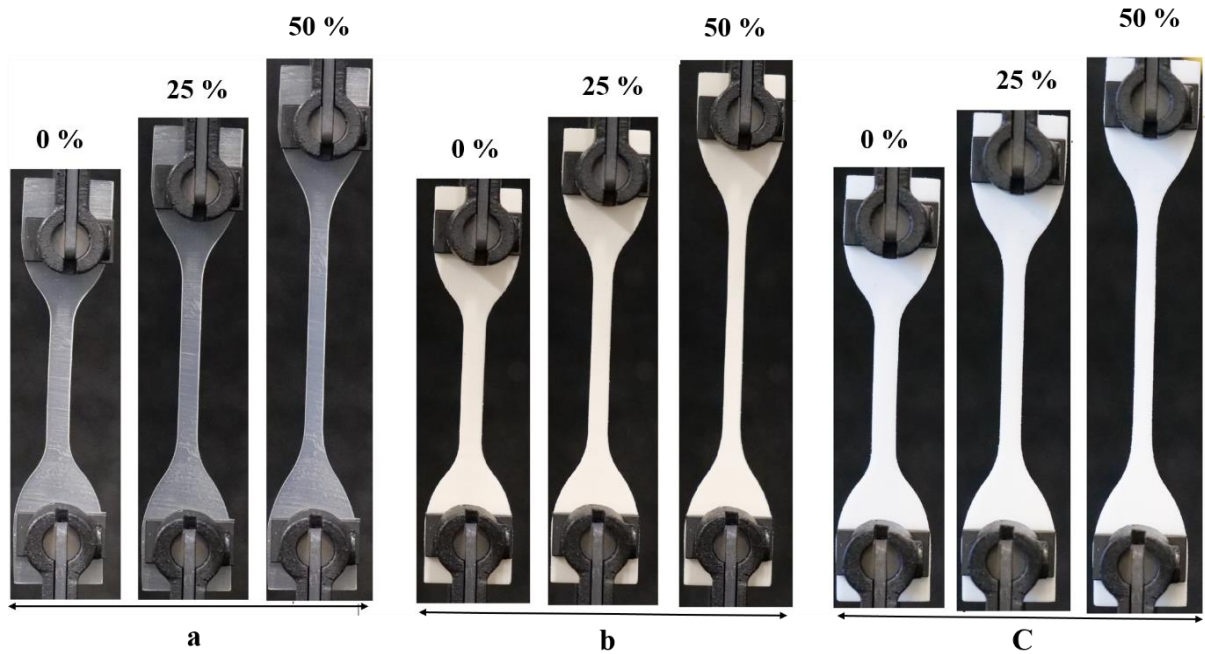

**Fig. 5** Samples of unfilled polyurethane (a), 551-40% (b) and 920-40% (c) at various strain levels (0%, 25% and 50% strain)

Photographs of unfilled polyurethane and SFs (551-40% and 920-40%) stretched to various strains are presented in Fig. 5 while the loading and unloading curves for both unfilled polyurethane and SF samples stretched to 25% strain are presented in Fig. 6. Here the SFs consisting of different HTMs (551 & 920) and volume fractions (2%, 10%, & 40%) are represented by HTM grade along with the volume fraction used. Neat polyurethane samples are termed as "unfilled" in this work. At the 25% strain level, the stress-strain curves of the unfilled samples follow a non-linear behaviour typically associated with soft polymer materials [25]. When the samples are unloaded at the same strain rate as the loading cycle, the unloading curves follow an almost identical path to that of the loading curves, indicating the almost perfectly elastic behaviour of the material at this strain rate when subjected to this level of strain. The hysteresis loop is barely visible during cyclic loading of unfilled samples. For the unfilled polyurethane, all five loading and unloading curves follow the same loading-unloading path as the first cycle, showing that the material returns to its original dimensions upon unloading. A similar behaviour has been reported by Meunier et al. [25] for unfilled rubber under cyclic tensile loading. For both 551 and 920 grade SFs with 2% and 10% volume fractions, the stress-strain curves also exhibit a non-linear response, similar to that of the matrix material, which reflects the dominance of the matrix material during extension at small to moderate microsphere volume fractions. However, for these volume fractions, a hysteresis curve is formed during cyclic loading-unloading associated with the energy dissipation of these materials. For SFs with 2% volume fraction, all five loading and unloading curves closely follow the same path as the first loading-unloading curve. However, for SFs with 10% volume fraction, after the first loading-unloading curve, the behaviour of successive loading-unloading curves starts to differ from the unfilled case and the SFs with 2% volume fraction. At 10% volume fraction, after the first loading cycle, the subsequent four loading curves deviate from the path of the first loading curve. These four subsequent successive loading curves follow a similar path however and a smaller force is needed for the same strain in the last four loading curves compared to the first loading curve. This behaviour can be associated with stress softening of the material during cyclic loading [26].

Both 551 and 920 grade SF samples with 40% volume fraction exhibit extremely different behaviour to the lower volume fraction samples, namely the emergence of an initial linear region followed by a non-linear stress-strain response. This small initial linear region can be attributed to an increase in the initial stiffness of the SFs with a higher volume fraction of microspheres. Similarly to 10% volume fraction SFs, stress-softening is also observed in the stress-strain curves of SFs with 40% volume fraction in the last four loading cycles. Interestingly, stress softening is even more prominent in 40% volume fraction samples as compared to the 10% volume fraction samples. Unlike the unfilled and 2% volume fraction, it is also observed that 10% and 40% volume fraction SFs do not return to their original configurations after the first loading cycle, exhibiting a residual strain. For SFs with 40% volume fraction, around 1-2% residual strain was recorded after tensile testing to 25% strain. The samples were left to recover for one week and it was noted that residual strain disappeared after this period.

After examining the initial cyclic response of unfilled and HTM SF samples to 25% strain, we subsequently studied the repeated cyclic curves to 25% strain after one week. The repeated cyclic testing results do not show any significant change compared to those for the virgin samples.

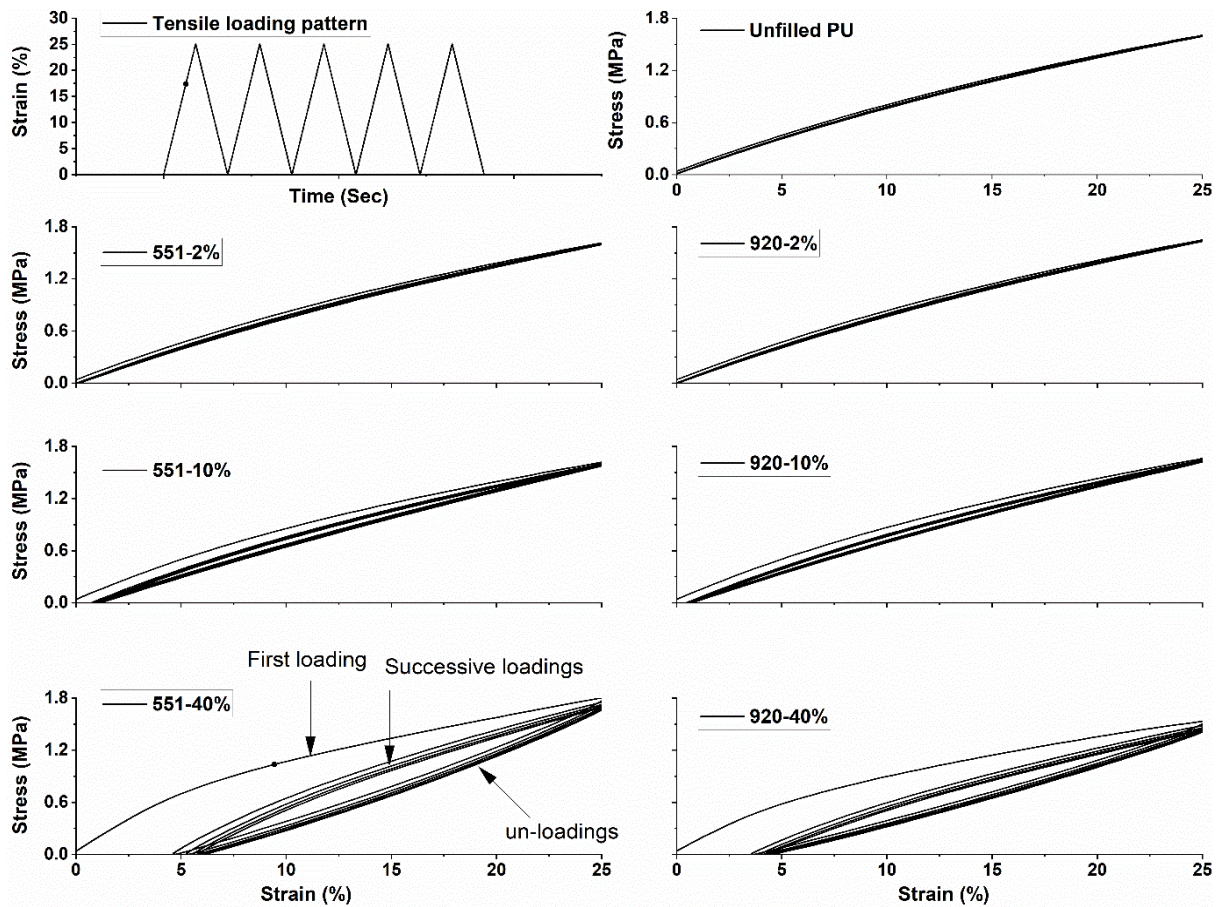

**Fig. 6** Stress-strain curves for unfilled, 551, and 920 HTM SFs up to 25% strain (virgin samples).

The repeated curves for the unfilled and HTM SFs are represented by (red) dotted lines in Fig. 7. Once again, the phenomenon of stress softening and residual strain is apparent, as is indicated in the stress-strain graphs for 10% and 40% volume fractions. We note the strong recoverability after initial testing, given that the repeat tests follow, almost exactly, the initial loading curves. After performing cyclic testing on virgin and pre-tested samples up to 25% strain, the same specimens are subsequently tested to 50% strain after waiting for a further week. The stress-strain response of the specimens up to 50% strain levels is presented in Fig. 8. The stress-strain curves for the unfilled and HTM SF samples exhibit similar trends to those observed in the testing of samples up to 25% strain. Here again, we observe a non-linear stress-strain behaviour for unfilled and HTM SFs with 2% and 10% volume fractions, and a small linear region for HTM SFs with 40% volume fraction. The phenomenon of stress softening and residual strain is visible once again for HTM SFs with 10% and 40% volume fractions and is particularly noticeable for the 40% case. After initial loading to 50% strain, the samples were left to relax for a week before proceeding to repeat the loading to the same level of strain. It is worth mentioning that samples of both 551 and 920 grades with 10% and 40% volume fractions did not return to their original configurations even after one week while exhibiting a residual strain of around 2-4%. The repeated test results for cyclic loading to 50% strain are presented in Fig. 9. The repeated stress-strain curves for unfilled and 2% volume fraction follow an almost identical path to the initially tested samples with an insignificant deviation in the loading-unloading curves. However, for HTM SFs with 10% and 40% volume fractions, the first loading curve of the repeat cycle does not follow the same path as the first loading curve of the initially tested samples. Stress-softening is observed for the first loading curve of these volume fractions (10% and 40%) under repeated loading. This phenomenon is especially profound for 40% volume fraction. The fact that stress softening is noticeable (particularly for the 10% and 40% SFs) indicates that the HTM

SFs have undergone permanent damage, perhaps associated with microsphere debonding which is supported by the presence of residual strain in 10% and 40% HTM SFs even after one week of loading to 50% strain. This therefore merits investigation of the damage and subsequent failure that HTM SFs undergo under tensile loading, which we report on in the next subsection.

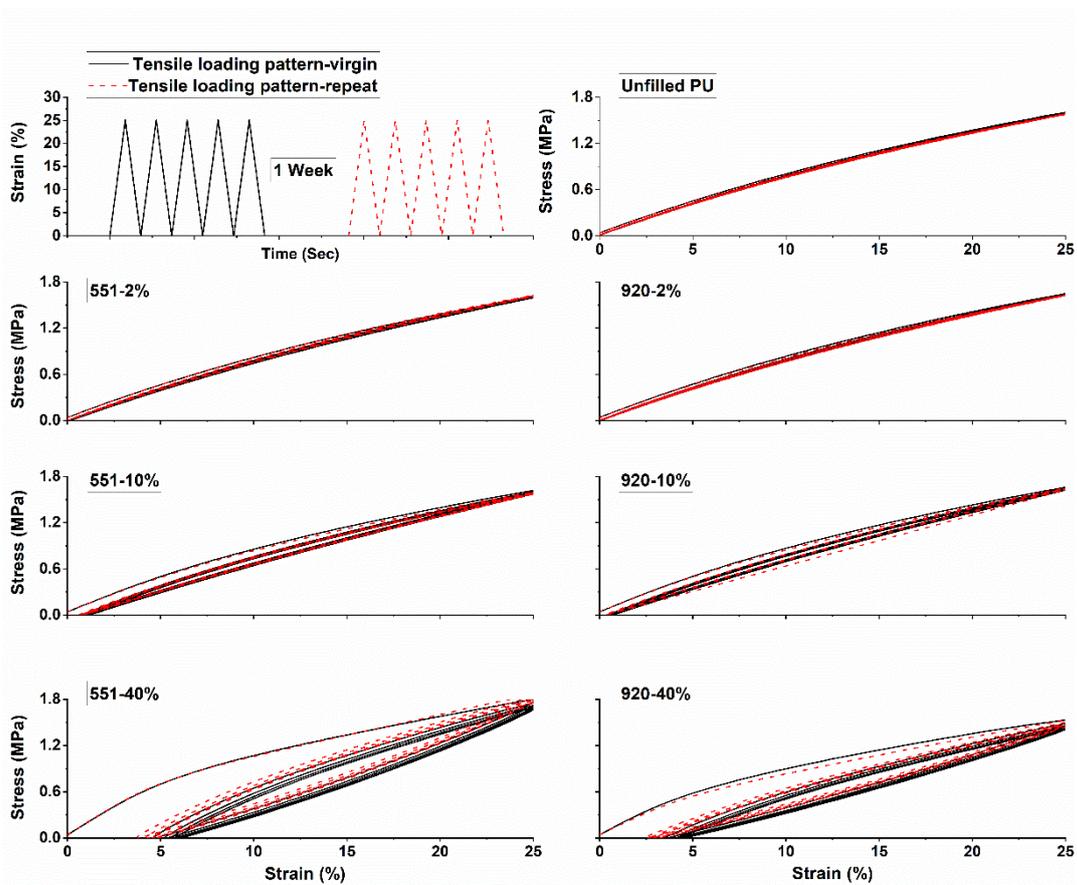

**Fig. 7** Representative stress-strain curves for unfilled PU and 551 and 920 HTM SFs up to 25% tensile strain. Initial testing cycles are represented by solid black lines and repeated testing cycles by dotted red lines. Strong recoverability after initial testing is highlighted, given that the repeat tests follow, almost exactly, the initial loading curves.

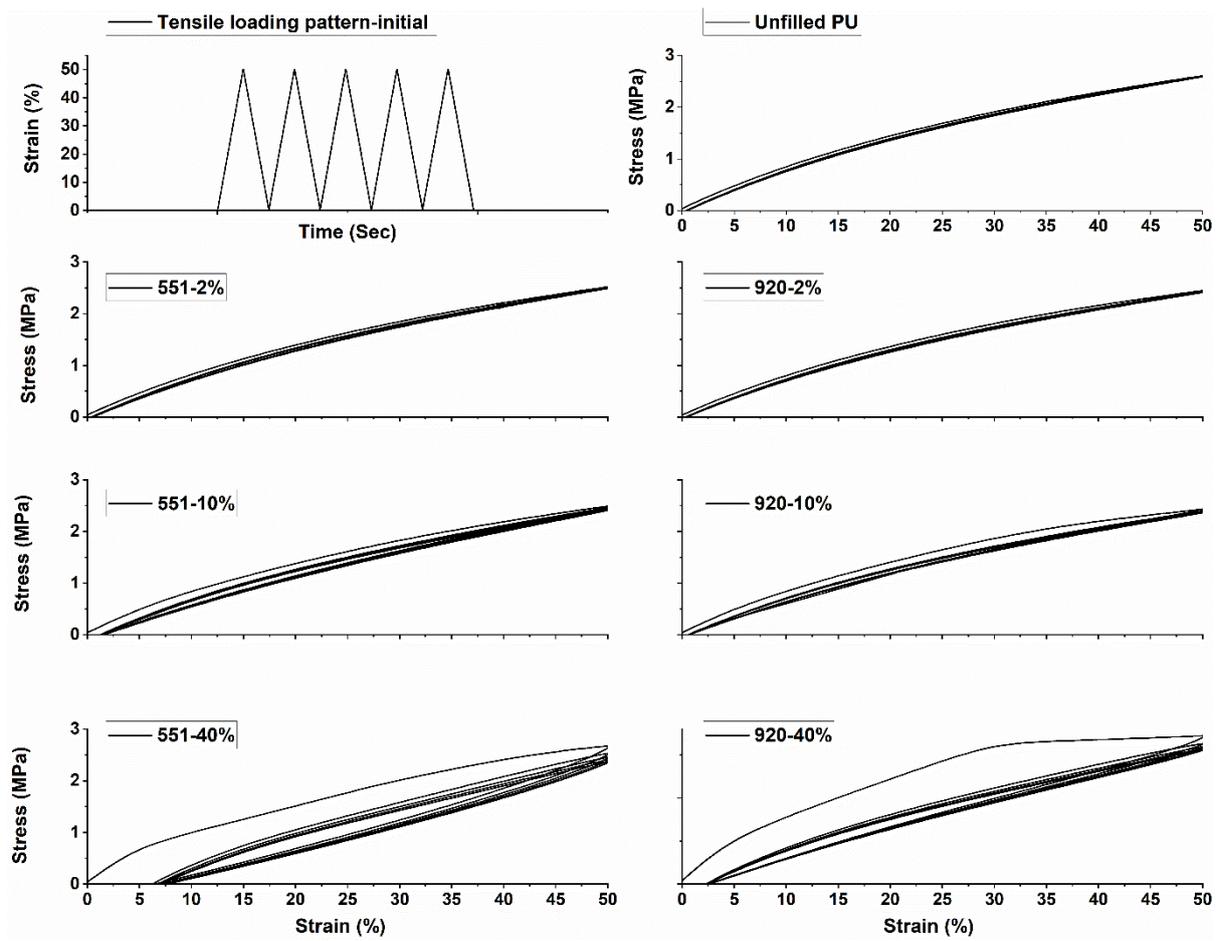

**Fig. 8** Stress-strain curves for unfilled PU and 551 and 920 HTM SFs up to 50% strain.

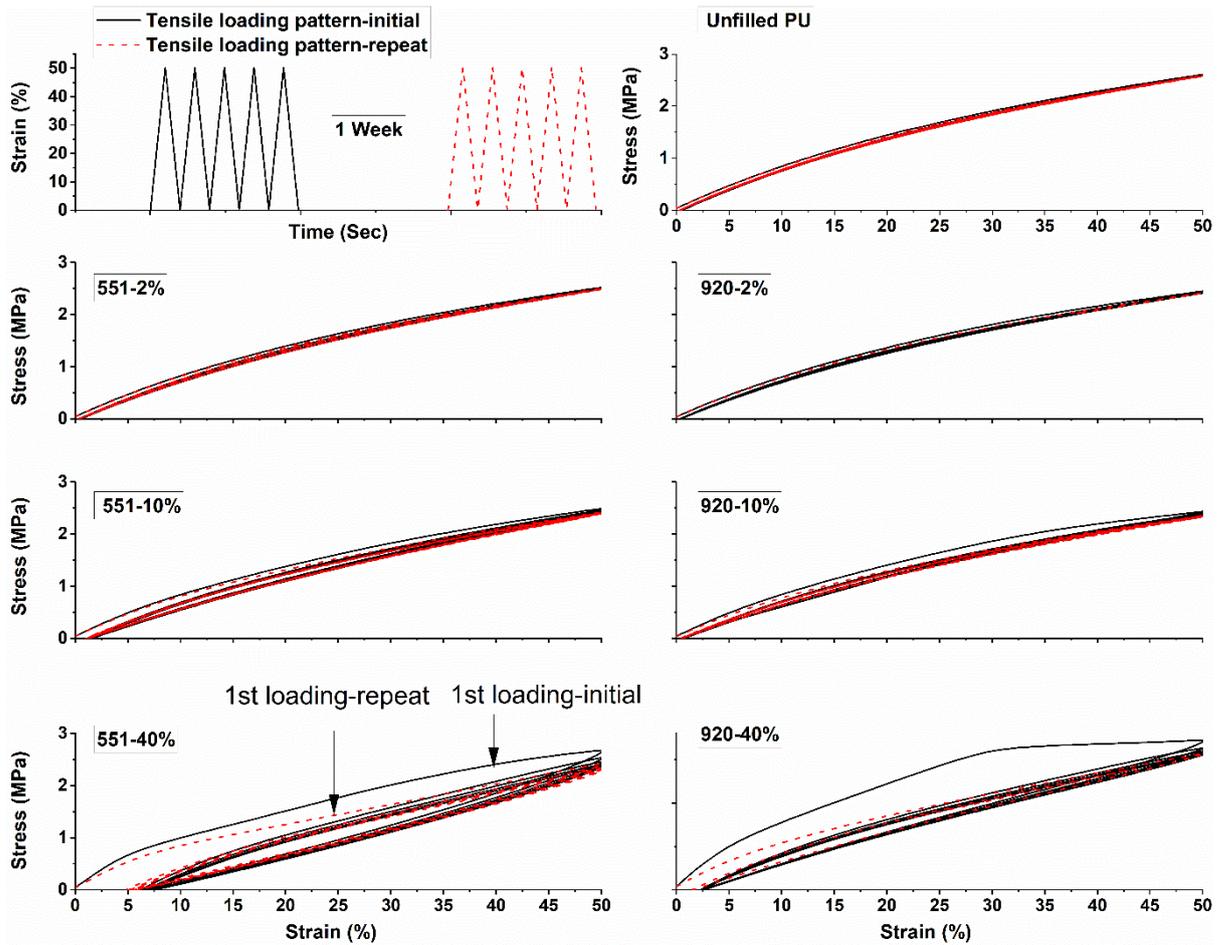

**Fig. 9** Representative stress-strain curves for unfilled PU and 551 and 920 HTM SFs up to 50% tensile strain. Initial testing cycles are represented by solid black lines and repeated testing cycles by dotted red lines.

3.2  Macroscopic deformation under monotonic loading

In order to assess the failure properties of the HTM SFs with the full range of microsphere volume fractions, the unfilled polyurethane and new, virgin HTM SF samples were subjected to monotonic tensile loading in order to reveal their tensile strength and elongation at break. The stress-strain results of the samples under monotonic tensile loading are presented in Fig. 10. Elongation at break is highest for unfilled polyurethane. Furthermore, elongation at break decreases with an increase in the microsphere volume fraction for both 551 and 920 SFs. At 40% volume fraction, the reduction in elongation at break of 551 and 920 SFs relative to the unfilled matrix is approximately 38% and 78%, respectively. The decrease in elongation at break can be attributed to higher stress concentrations at the matrix-microsphere interfaces resulting in matrix-microsphere debonding [3]. Higher volume fractions of microspheres allow more microspheres to debond from the matrix. A similar behaviour has been reported in the literature where a decrease in elongation at break of polymeric composites was observed with an increase in the filler content [27, 28]. Also, elongation at break of 920 HTM SFs is lower than 551 SFs. In particular, at 40% volume fraction, elongation at break of 920 HTM SFs is 65% lower than 551 HTM SFs. This can be attributed to the larger diameter of 920 microspheres, resulting in higher interfacial area of individual microspheres and consequently a lower elongation at break.

We note that the tensile strength of HTM SFs is also lower than that of unfilled polyurethane. In fact the tensile strength of 551 and 920 HTM SFs is almost 65% and 85% lower than unfilled polyurethane respectively, at breaking point. It is worth noting that for 920 HTM SFs, in comparison to cyclic loading (where as we showed above, there is evidence of permanent damage at 50% strain), the elongation at break under monotonic loading is 62% indicating that fatigue loading results in a lower breaking strain. Furthermore, following a similar trend to elongation at break, tensile strength decreases with increase in microsphere volume fraction. Yu et al. [29] observed a similar behaviour for the tensile strength of SFs manufactured with ceramic microspheres. It has been reported that the tensile behaviour of filled elastomers is highly dependent on the bond strength between the filled particles and the polymer matrix [30]. For a very weak bond strength, the filler-polymer bond fails immediately upon straining, creating small voids next to the filler particle. Due to the absence of physical reinforcement, the material is weak and highly extensible. However, when the bond between the filler and polymer is strong, the filled material exhibits a relatively strong modulus with considerable reduction in elongation at break. In the present case, as the filled material is relatively stiff with lower elongation at break than unfilled polyurethane, we can interpret the bond between filler and polyurethane rubber as being relatively strong.

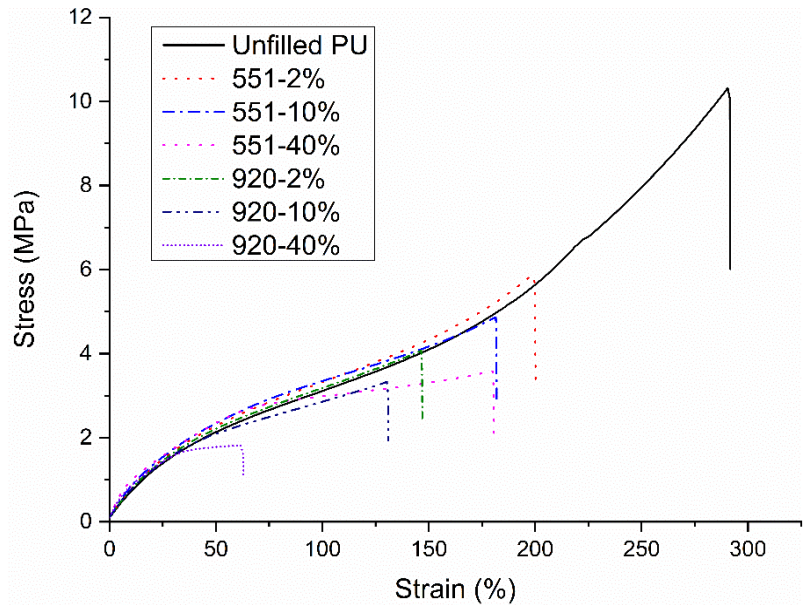

**Fig. 10** Stress-strain curves of unfilled PU and HTM SFs under monotonic tensile loading.

3.3  Modulus of unfilled polyurethane and SFs

During tensile testing, as should be expected, it is observed that the small strain stiffness of HTM SFs is higher than unfilled polyurethane (Fig. 11). That is, the elastic modulus (Young's modulus) increased with increasing volume fractions of microspheres in the polyurethane matrix. Referring to Fig. 11, at 40% volume fraction, the elastic modulus of 551 and 920 SFs are 27% and 19% higher than unfilled polyurethane, respectively. Gupta et al. [7] observed an increase in Young's modulus of several compositions of HGM SFs. Other studies reported a decrease in Young's modulus of SFs upon an increase in microsphere volume fraction [18, 31]. However, they stated that the SFs comprise a weaker filler compared to the matrix material, and so inclusion of weaker particles in the stronger matrix decreases the Young's modulus of the SF. It should also be noted that the elastic modulus of the 551 grade HTM SF is higher than the 920 grade. For example, at 40% volume fraction, the elastic modulus of HTM SFs comprising 551 microspheres is 6.5% higher than for SFs containing 920 microspheres. This can be attributed to larger wall-thickness-to-radius ratio, as it has been reported that SFs containing microspheres of the same material with larger wall thickness-to-diameter ratio generally exhibit improved mechanical properties [3]. In addition to Young's modulus, the specific modulus is also calculated in the present study (Fig. 11).

The specific modulus follows an analogous trend to the Young's modulus, given the relatively simple mixture theory variation in the effective density of the SFs reported above. At 40% volume fraction, the specific modulus of 551 and 920 HTM SFs are 107% and 93% higher than unfilled polyurethane respectively, which are significant increases for applications where lightweight stiff materials are required. A similar trend of specific tensile modulus of SFs with increasing volume fraction of microspheres has been reported in the literature for HGM SFs [7]. Wouterson et al. [3] observed an increase in specific modulus of HGM SFs compared to neat epoxy resin upon the inclusion of a small amount of glass microspheres, while for SFs manufactured with a higher volume fraction of microspheres they reported a decrease in specific modulus. They infer that the relative reduction in strength of HGM SFs with higher volume fractions of microspheres is larger than the relative reduction in density.

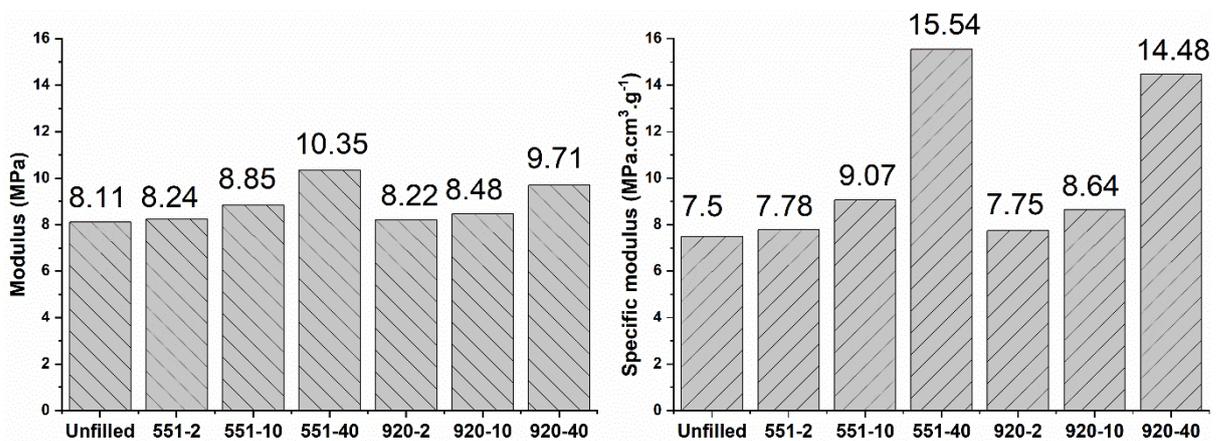

**Fig. 11** Elastic modulus (left) and specific modulus (right) of unfilled PU and SFs.

## 4. Modelling the elastic properties of SFs

4.1  Overview

We now employ the results above from our tensile testing in order to address the physical characterization of HTMs, using a mathematical model, combining geometrical polydispersity data obtained via imaging techniques [24] with the measured tensile response of SFs, as reported in Section 3. The model's stiffness prediction for a SF containing microspheres with sizes and aspect ratios distributed according to the data obtained from imaging allows us to solve iteratively for the best fitting values of the unknown (or unverified) properties of the microspheres. In particular, we extend characterization of HTMs in [24] by computing values for their Young's moduli and Poisson's ratios when under tension, comparing these to analogous values found in [24] via fits to compression experiments [4], and we determine a representative value of the shell thickness of 551 DE 40 d42 grade HTMs, which until now has not been reported.

4.2  Details of the mathematical model

The model used here to calculate the effective elastic properties of a HTM SF, under the assumptions of linear elasticity (small strain), is based on an extension of the Generalized Self-Consistent Method (GSCM) [24]. This method was developed to represent composites containing solid monodisperse cylindrical or spherical particles [32, 33], and it was later extended to handle multi-layered shell-like inclusions with internal voids [34]. A key development was the combination of the GSCM with a volume-averaging approach [35], to enable the modelling of SFs containing polydisperse inclusions [2], in particular HTMs of different diameters and/or shell thicknesses, and potentially different elastic properties. This versatility, and a strong performance in comparison to alternative methods [36], are important advantages of the extended GSCM for our purposes of accurately modelling specified distributions of microspheres within a syntactic foam.

Suppose that a syntactic foam contains $J$ distinct types of microsphere, with volume fractions $\{c_j\}$ where $j \in \{1, 2, \ldots, J\}$ such that $\sum_1^J c_j = c_0$, where $c_0$ is the overall volume fraction of microspheres in the SF. Adapting the original GSCM [32] for the case of hollow spherical inclusions of outer radius $a_j$ and shell thickness $h_j$, a four-phase problem is solved in which a single, isolated three-phase composite sphere of type $j$ is surrounded by an infinite "equivalent homogeneous medium" phase. The geometry of a single composite sphere is shown in Fig. 12. The void phase occupies the domain $0 \leq r \leq a_j - h_j$, the shell phase (with shear modulus $\mu_{s,j}$ and bulk modulus $\kappa_{s,j}$) occupies $a_j - h_j \leq r \leq a_j$, the matrix phase ($\mu_{m,j}$, $\kappa_{m,j}$) occupies $a_j \leq r \leq b_j$, and the equivalent homogeneous medium phase ($\mu_{e,j}$, $\kappa_{e,j}$) (i.e. the effective medium) occupies $r \geq b_j$. The shell aspect ratio of type $j$ is denoted as $\eta_j = 1 - h_j/a_j$, and the composite sphere's radius $b_j$ is prescribed by imposing that the volume fraction of a single inclusion within its own composite sphere is equal to $c_0$, i.e. $c_0 = (a_j/b_j)^3$. A pure void fraction can be represented by a type with $h_j = 0$ (assuming spherical voids); the incorporation of unwanted voids in modelling was found to be important in [2], but due to the closer similarity of the measured and theoretical densities of the foam samples in this study (see Table 2) we do not utilize a pure void fraction in our implementation here.

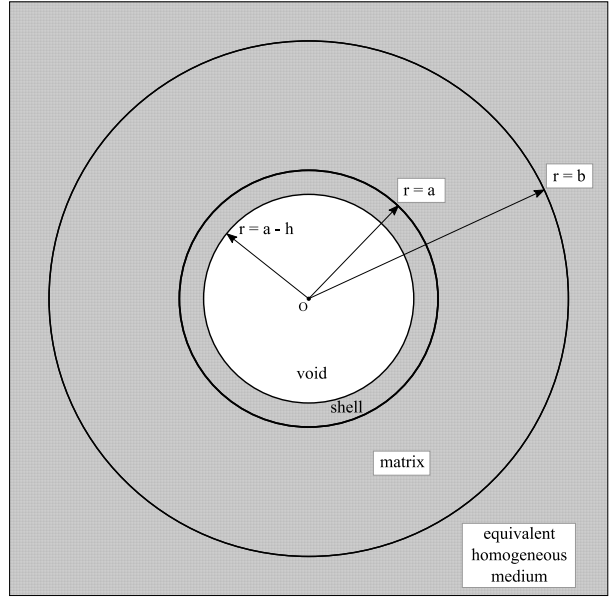

**Fig. 12** The geometry of a single multiphase composite sphere in the model.

Equations for the shear and bulk moduli of the equivalent homogeneous phase are derived (separately) by solving linear systems arising from the sets of boundary conditions for the displacement and stress fields when the four-phase problem is subjected to shear and hydrostatic far-field conditions respectively. Our derivation differs slightly from that in [2] because we solve for the displacement fields within each composite sphere under the same far-field conditions as in [32], although the resulting systems of equations are equivalent.

In the original monodisperse GSCM, a strain energy condition of equivalence to the case with no inclusion is used to impose the self-consistency of the model [32]. In the polydisperse extension of the GSCM [2], the representative volume element (RVE) is an assembly of the composite spheres associated with the $J$ types of microsphere, such that the proportion by volume of composite spheres of type $j$ in the RVE is $c_j/c_0$. The self-consistency of the polydisperse model is imposed by volume averaging the stress and strain fields over the RVE, defining effective shear and bulk moduli ($\mu^*, \kappa^*$) of the RVE as appropriate ratios of averaged stress and strain components, and then setting $\mu_{e,j} = \mu^*$ and $\kappa_{e,j} = \kappa^*$ for each $j$.

This results in the following pair of equations for $\mu^*$ and $\kappa^*$, which are coupled in the polydisperse case (i.e. $J > 1$):

$$\mu^* = \frac{\frac{1}{V}\int_V \mu \cdot (e_{xx} - e_{yy})dV}{\frac{1}{V}\int_V (e_{xx} - e_{yy})dV} = \frac{\sum_1^J \frac{c_j}{c_0} \cdot \frac{1}{V_j}\int_{V_j} \mu \cdot (e_{xx} - e_{yy})dV}{\sum_1^J \frac{c_j}{c_0} \cdot \frac{1}{V_j}\int_{V_j} (e_{xx} - e_{yy})dV},$$

$$\kappa^* = \frac{\frac{1}{V}\int_V \kappa \, \mathrm{tr}\, \mathbf{e}\, dV}{\frac{1}{V}\int_V \mathrm{tr}\, \mathbf{e}\, dV} = \frac{\sum_1^J \frac{c_j}{c_0} \cdot \frac{1}{V_j}\int_{V_j} \kappa \, \mathrm{tr}\, \mathbf{e}\, dV}{\sum_1^J \frac{c_j}{c_0} \cdot \frac{1}{V_j}\int_{V_j} \mathrm{tr}\, \mathbf{e}\, dV},$$

where **e** is the strain tensor that satisfies the four-phase linear elasticity problem under the relevant far-field conditions (shear for the $\mu^*$ equation, hydrostatic for the $\kappa^*$ equation), $\mu$ and $\kappa$ are the spatially varying shear and bulk moduli (piecewise constant in each composite sphere), $V$ is the volume of the RVE, and $V_j$ is the volume of a composite sphere of type $j$. For each set of input parameters, these implicit equations are solved numerically in MATLAB R2019a via "fsolve" to output the effective shear and bulk moduli (and hence the Young's modulus and Poisson's ratio) of a particular SF.

4.3   Input parameters

For each calculation of the effective elastic properties of a particular SF, using the model, we specify as inputs the linear elastic properties of both the matrix material and the microsphere shell material, in addition to the geometric polydispersity distribution of microspheres (different for each HTM grade), and the overall volume fraction of microspheres in the SF. The matrix material itself (i.e. unfilled polyurethane) is taken to have a tensile Young's modulus of 8.11 MPa and a Poisson's ratio of 0.49 [37]; these do not vary throughout this study.

Representative figures for the elastic properties of the HTM shell material were obtained in [24], based on a comparison of model outputs (via the same model employed here) with previous experimental measurements of the response of samples under uniaxial compression [4]. We therefore use a shell Young's modulus of 1.67518 GPa and a shell Poisson's ratio of 0.201136 in our initial calculations here, but we also allow these parameters to vary in order to assess the extent to which the appropriate characterization of the microspheres differs in the context of SFs under tension, as opposed to compression.

We assume that the shell material of every HTM has the same elastic properties, independent of diameter and thickness, and we also assume that the geometric polydispersity distribution for each grade of HTM is independent of the overall volume fraction of microspheres in the SF.

In Table 1 we provide values of a representative average diameter and shell thickness for each grade, quoting manufacturer specifications [38, 39] and volume-weighted mean diameters obtained as statistics of the full distributions in [24].

In this study we use the model to calculate the effective elastic properties of SFs containing monodisperse microspheres, using data from Table 1 as inputs, and those of SFs containing polydisperse microspheres, using discretizations of the full geometric distributions of the two HTM grades. The polydispersity in microsphere aspect ratio is controlled via the diameter distribution alone, as the shell thickness of the 920 grade was found in [24] to be approximately constant at 290 nm independent of diameter, while the thickness of the 551 grade was not reported. We assume here that the latter thickness is also approximately constant. In the model, the diameter distribution of each grade is discretized using 5 μm sub-intervals, such that all microspheres with diameters lying in a given sub-interval are represented by the volume-weighted average diameter of that sub-interval.

For both 920 and 551 grades, we compare the model's predictions of the Young's modulus of HTM SFs of different volume fractions to our experimental measurements of the tensile response of samples (as in Fig. 11), using an iterative process to determine the best fitting values of the microspheres' elastic properties (920 grade) and their representative shell thickness (551 grade). Given the assumptions inherent in the GSCM, in particular that predictions for highly filled composites are likely to be less accurate, we restrict our fitting criteria to the mean squared error in the Young's modulus of the 2% and 10% volume fraction samples only, excluding the 40% case.

4.4  920 grade

In Fig. 13(a) we show the Young's modulus of polyurethane SFs containing 920 grade HTMs, as a function of volume fraction. Our experimental data are plotted alongside model predictions for three different cases. The green (dotted) line shows the model calculation for a monodisperse case using the manufacturer specifications for microsphere diameter and shell thickness (as in Table 1), together with a shell Young's modulus of 1.67518 GPa and a shell Poisson's ratio of 0.201136 as obtained in [24]. The black (dashed) line is the model calculation for the polydisperse case with the same elastic properties of the shells. Comparing these two lines to the experimental data, it is clear that an improved representation of the SF's response is obtained in the model by incorporating the full distribution of microsphere diameters obtained by imaging techniques, as well as the more accurate estimate of the 920 grade shell thickness.

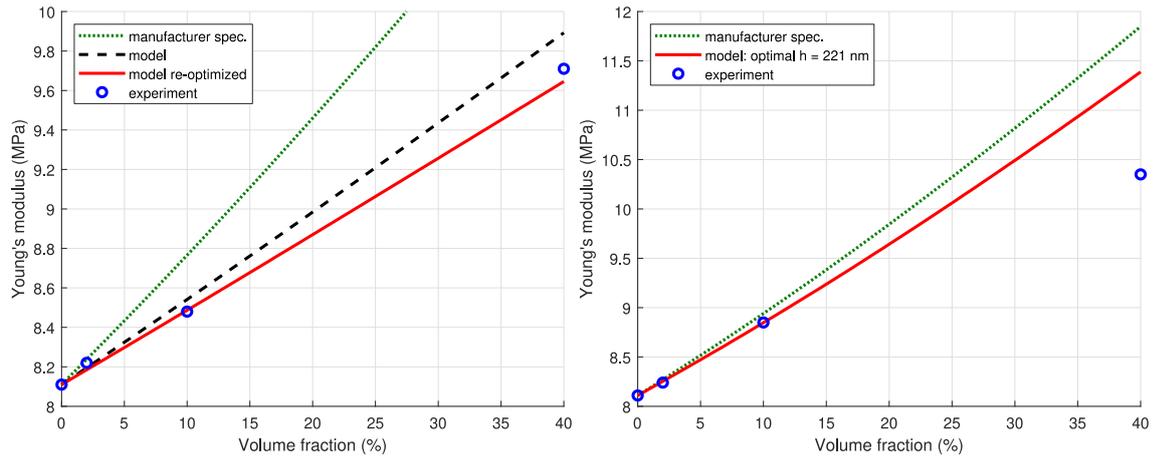

Fig. 13 (a & b) - Young's modulus of SFs containing HTMs; 13(a) (left) - 920 grade, 13(b) (right) - 551 grade. Dotted green lines: monodisperse model predictions using manufacturer specifications of microsphere geometry. Dashed black line: polydisperse model prediction using microsphere properties (geometric and elastic) from [24]. Solid red lines: polydisperse model predictions using microsphere geometry from [24] and re-optimized elastic properties; in (b) a best fitting value of the shell thickness is used. Blue circles: experimentally determined tensile response of samples.

In the third case plotted in Fig. 13(a), we allow the elastic properties of the shell to vary in order to obtain the best fit of the polydisperse model to the experimental data (2% and 10% volume fractions only), thereby re-characterizing the HTMs' elastic properties within a SF under tension. We find that the best fit is attained via a shell Young's modulus of 1.56194 GPa (within bounds of 0.5 to 2.5 GPa) and a shell Poisson's ratio of 0.202989 (within bounds of 0.2 to 0.3), and this optimal case is shown as the red (solid) line in Figure 13(a). In a fourth case (not plotted) we find also an optimal fit for a monodisperse SF with microsphere diameter equal to the measured volume-weighted mean diameter of 920 grade HTMs (as given in Table 1); the corresponding shell Young's modulus is 1.66447 GPa, and the shell Poisson's ratio is 0.200011, within the same bounds. This fourth case produces a model prediction which is practically identical to the red (solid) line in Figure 13(a). From this we deduce that it is possible to obtain a good monodisperse model approximation to the elastic properties of a SF containing polydisperse HTMs, provided that the representative monodisperse microsphere has the most appropriate available geometric properties, i.e. those based on the relevant statistics of the full polydisperse distribution.

4.5   551 grade

In Fig. 13(b) we plot the Young's modulus of polyurethane SFs containing 551 grade HTMs, as a function of volume fraction. Our experimental data are plotted alongside model predictions for two different cases. The green (dotted) line shows the model calculation for a monodisperse case using the manufacturer specifications for microsphere diameter and shell thickness (as in Table 1), together with a shell Young's modulus of 1.66447 GPa and a shell Poisson's ratio of 0.200011; these elastic properties are set equal to those of the best fitting monodisperse case for 920 grade HTMs, as found in section 4.4.

For the 551 grade, the manufacturer's estimate of the average microsphere diameter is almost equal to the volume-weighted mean diameter obtained via imaging (see Table 1) - this is not the case for the 920 grade. Consequently, the monodisperse model prediction using the manufacturer specifications for the 551 grade shows a considerably better agreement with the experimental data, compared to the 920 grade. However, the shell thickness of the 551 grade has not been determined via imaging, so we employ both the monodisperse and polydisperse models to find the best fitting value of this parameter, and thus complete our characterization of 551 grade HTMs.

In the second case plotted in Figure 13(b), we allow the shell thickness (but not the diameter distribution) of the 551 grade to vary in order to obtain the best fit of the polydisperse model to the experimental data (2% and 10% volume fractions). Using a shell Young's modulus of 1.56194 GPa and a shell Poisson's ratio of 0.202989 (the elastic properties of the best fitting polydisperse case for 920 grade HTMs, as found in section 4.4), we find that the best fit is attained via a 551 grade shell thickness of 0.221043 μm (within bounds of 0.1 to 0.5 μm), and this optimal case is shown as the red (solid) line in Fig. 13(b). In a third case (not plotted) we find also an optimal fit for the monodisperse model with microsphere diameter equal to the measured volume-weighted mean diameter of the 551 grade HTMs (as given in Table 1), and with a shell Young's modulus of 1.66447 GPa and a shell Poisson's ratio of 0.200011 (the elastic properties of the best fitting monodisperse case for 920 grade HTMs); the corresponding best fit is attained via a 551 grade shell thickness of 0.220051 μm, within the same bounds. This third case produces a model prediction which is practically identical to the red (solid) line in Fig. 13(b). The similarity of the 551 grade shell thicknesses estimated via both the monodisperse and polydisperse models (220 nm and 221 nm respectively) indicates that our method of characterizing microsphere properties is not critically dependent upon access to a high resolution microsphere diameter distribution, provided that an accurate representative mean diameter is available.

## 5.  Thermogravimetric analysis (TGA)

The thermal degradation of unfilled polyurethane and HTM SFs is examined by using thermogravimetric analysis (TGA). For this purpose, TGA Q500 from TA Instruments was employed under a nitrogen atmosphere. The weight of the specimens was around 10 mg. The samples were heated from room temperature to 600 °C at a heating rate of 10 °C/min. TGA curves for unfilled polyurethane and HTM SFs are presented in Fig. 14. The curves for all configurations are very similar. Only a slight decrease in temperatures at 5% and 10% weight loss is recorded with addition of HTMs (Table 3). Similar results have been reported for SFs manufactured from silicone-modified epoxy and HTMs [40]. The authors have previously reported insignificant difference in glass transition temperature (Tg) of unfilled polyurethane and HTM SFs [37] which show HTMs have very little influence on thermal characteristics (TGA and Tg) of syntactic foam composites.

**Table 3** Thermal properties of unfilled polyurethane and HTM SFs.

| Sample | Unfilled PU | 551-2% | 551-10% | 551-40% | 920-2% | 92-10% | 920-40% |
|---|---|---|---|---|---|---|---|
| Temp. (°C) at 5% weight loss | 319 | 318 | 312 | 303 | 317 | 314 | 308 |
| Temp. (°C) at 10% weight loss | 333 | 331 | 327 | 319 | 331 | 328 | 322 |

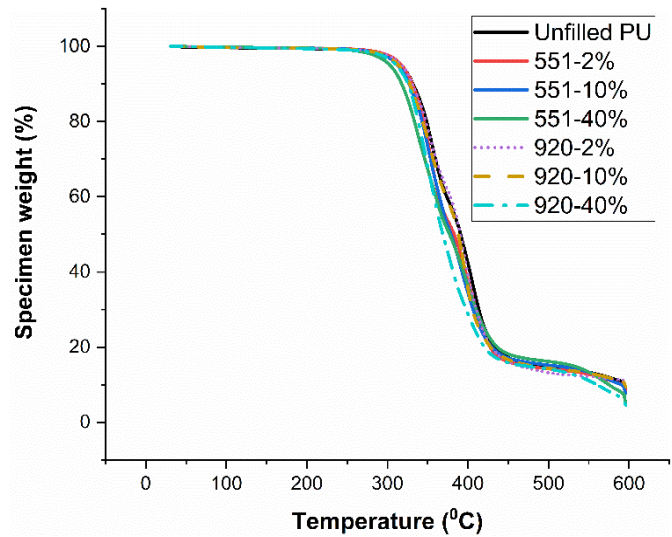

**Fig. 14** TGA curves of the unfilled polyurethane and HTM SFs.

## 6. Conclusions

In this work we have studied the tensile properties of unfilled polyurethane and HTM SFs under cyclic and monotonic loadings. The HTM SFs were manufactured by adding HTMs of two different grades (551 and 920), with distributions of mean-wall thicknesses and diameters, embedded inside a polyurethane matrix in various volume fractions. To investigate cyclic tensile properties, five loading-unloading cycles were performed on each specimen up to strains of 25% and 50%. The cyclic tensile testing on each sample was conducted twice. The objective of the secondary testing (performed after one week of initial testing) was to reveal the time dependent stress-strain response of these HTM SFs. To understand the damage and elongation at break, monotonic loading was performed to failure. Furthermore, by employing optimization techniques, linear elastic properties of the microspheres and an average shell thickness of the 551 grade were inferred by comparing experimental results to predictions from the Generalized Self-Consistent Method, incorporating polydispersity data on the size distribution of the microspheres. Finally we determined thermal degradation of unfilled polyurethane and HTM SFs by employing thermogravimetric analysis. The following key conclusions can be drawn from this study:

1- During cyclic loading, the stress-strain curves for unfilled polyurethane and SFs with 2% microspheres presented non–linear behaviour while an initial linear region emerged for SFs with 10% and 40% volume fractions, which was due to an increase in the initial stiffness of HTM SFs with higher volume fractions of microspheres.

2- Loading-unloading curves for unfilled polyurethane and HTM SFs with 2% volume fraction followed a similar pattern showing the elastic nature of these materials, while a hysteresis loop appeared for SFs with 10% and 40% volume fractions depicting energy dissipation during cyclic loading for HTM SFs with higher volume fractions of microspheres.

3- Stress softening and residual strains were recorded for SFs with 10% and 40% volume fraction. The residual strain disappeared when the samples were relaxed after 25% strain whilst it remained present after 50% strain, which we ascribe to permanent damage (microsphere debonding from matrix material) under cyclic loading at this strain level.

4- Monotonic loading to failure showed a decrease in elongation at break of HTM SFs with increase in microsphere volume fraction. We speculate that high stress concentrations at the microsphere-matrix interface resulted in debonding of microspheres from the matrix material. With higher microsphere volume fractions, the interfacial area increased, resulting in a decrease in elongation at break. Elongation at break was lowest for HTM SFs 920 grade with 40% volume fraction, which was due to the fact that microspheres of 920 grade have a larger mean diameter than those of 551 grade and consequently a greater interfacial area, which resulted in the lowest elongation at break.

5- Both the Young's modulus and the specific modulus increased with increase in microsphere volume fractions. The increase in specific modulus was significantly higher at 40% volume fraction for both 551 and 920 grade HTM SFs (107% and 93% higher than unfilled polyurethane respectively) revealing that HTM SFs can be good candidates for applications where lightweight stiff materials are required.

6- Previous estimates of predicted shell properties gave a good fit to the experimental data. Using the model, the 551 grade shell thickness, previously unknown, was predicted to be approx. 220 nm (monodisperse) and 221 nm (polydisperse). The similarity of these two predictions indicates that characterizing microsphere properties is not critically dependent on the full polydisperse distribution, provided an accurate representative mean diameter is known. Interestingly, in the optimization procedure, if all shell properties are left free then there are multiple solutions in the parameter space such that thin stiff shells give similar results to thick soft shells. Similar optimization tools could be employed in the future in order to predict other properties of the shell, e.g. thermal and electrical properties, given macroscopic measurements.

7- The thermogravimetric analysis of unfilled polyurethane and HTM SFs revealed an insignificant effect of HTMs on the thermal degradation of HTM SFs.

Characterising the behaviour of new HTM SFs under different loading states is important in order to understand their potential in a variety of future applications. In particular when considering the loading type to which a component may be subjected, it is critical to be able to predict if a sample will fail, if it can successfully absorb energy under cyclic loading, and in its simplest state, what its initial linear elastic Young's modulus is under tension (and how this compares to its modulus under compression). As we have seen, combining experimental loading data with modelling techniques also permits more detailed microstructural characterisation, such as shell thickness predictions. Such analysis is potentially extremely powerful and future experiments will investigate the efficacy of such methods for the 551 grade provided here. Results obtained have the potential to prove very powerful in terms of the future design of lightweight composites in a broad range of applications such as those mentioned in the introduction of this paper.

# Acknowledgements

We acknowledge Alison Daniel (Thales UK) for sample manufacturing. The authors are grateful to the Engineering and Physical Sciences Research Council (EPSRC) for funding via grants EP/L018039/1 and EP/S019804/1.
# References

[1] Gupta N, Zeltmann SE, Shunmugasamy VC, Pinisetty D. Applications of Polymer Matrix Syntactic Foams. JOM. 2014;66:245-54.
[2] Bardella L, Genna F. On the elastic behavior of syntactic foams. International Journal of Solids and Structures. 2001;38:7235-60.
[3] Wouterson EM, Boey FYC, Hu X, Wong S-C. Specific properties and fracture toughness of syntactic foam: Effect of foam microstructures. Composites Science and Technology. 2005;65:1840-50.
[4] Yousaf Z, Smith M, Potluri P, Parnell W. Compression properties of polymeric syntactic foam composites under cyclic loading. Composites Part B: Engineering. 2020;186:107764.
[5] Gupta N, Ricci W. Comparison of compressive properties of layered syntactic foams having gradient in microballoon volume fraction and wall thickness. Materials Science and Engineering: A. 2006;427:331-42.
[6] Gupta N, Woldesenbet E, Mensah P. Compression properties of syntactic foams: effect of cenosphere radius ratio and specimen aspect ratio. Composites Part A: Applied Science and Manufacturing. 2004;35:103-11.
[7] Gupta N, Ye R, Porfiri M. Comparison of tensile and compressive characteristics of vinyl ester/glass microballoon syntactic foams. Composites Part B: Engineering. 2010;41:236-45.
[8] Gupta N, Woldesenbet E. Microballoon Wall Thickness Effects on Properties of Syntactic Foams. Journal of Cellular Plastics. 2004;40:461-80.
[9] Porfiri M, Gupta N. Effect of volume fraction and wall thickness on the elastic properties of hollow particle filled composites. Composites Part B: Engineering. 2009;40:166-73.
[10] Cochran JK. Ceramic hollow spheres and their applications. Current Opinion in Solid State and Materials Science. 1998;3:474-9.
[11] d'Almeida JRM. An analysis of the effect of the diameters of glass microspheres on the mechanical behavior of glass-microsphere/epoxy-matrix composites. Composites Science and Technology. 1999;59:2087-91.
[12] Orbulov IN, Ginsztler J. Compressive characteristics of metal matrix syntactic foams. Composites Part A: Applied Science and Manufacturing. 2012;43:553-61.
[13] Saha MC, Mahfuz H, Chakravarty UK, Uddin M, Kabir ME, Jeelani S. Effect of density, microstructure, and strain rate on compression behavior of polymeric foams. Materials Science and Engineering: A. 2005;406:328-36.
[14] Kim HS, Plubrai P. Manufacturing and failure mechanisms of syntactic foam under compression. Composites Part A: Applied Science and Manufacturing. 2004;35:1009-15.
[15] Everett RK, Matic P, Harvey Ii DP, Kee A. The microstructure and mechanical response of porous polymers. Materials Science and Engineering: A. 1998;249:7-13.
29